\documentclass[aps,pre,twocolumn]{revtex4-1}
\usepackage{graphicx}
\usepackage{hyperref}
\usepackage{xcolor}
\usepackage{amsmath}
\usepackage{amsfonts}
\usepackage{amssymb}
\usepackage{varioref}
\usepackage{float}
\usepackage{dcolumn}   % needed for some tables
\usepackage{subfigure}
\begin{document}
\title{Beating to rotational transition of a clamped active ribbon-like filament}
\author{Shalabh K. Anand \textit{$^{a}$}}
\email{skanand@iiserb.ac.in}
%\affiliation{Department of Physics,\\ Indian Institute of Science Education and Research, \\Bhopal 462 066, Madhya Pradesh, India}

\author{Raghunath Chelakkot \textit{$^{b}$}}
\email{raghu@phy.iitb.ac.in}
%\affiliation{\textit{$^b$}~Department of Physics, Indian Institute of Technology Bombay Powai, Mumbai 400 076, Maharashtra, India}

\author{Sunil P. Singh \textit{$^{a}$}}
\email{spsingh@iiserb.ac.in}
%\noaffiliation
\affiliation{\textit{$^a$}~Department of Physics,\\ Indian Institute of Science Education and Research, \\Bhopal 462 066, Madhya Pradesh, India \\ \textit{$^b$}~Department of Physics, Indian Institute of Technology Bombay Powai, Mumbai 400 076, Maharashtra, India}
%\date{\today}

\begin{abstract}
We present a detailed study of a clamped ribbon-like filament under a  compressive active force using Brownian dynamics simulations. We show that a clamped ribbon-like  filament is able to capture beating  as well as rotational motion under the compressive force. The nature of oscillation is governed by the torsional rigidity of the filament. The frequency of oscillation is almost independent of the torsional rigidity. The beating of the filament gives butterfly shape trajectory of the free-end monomer, whereas rotational motion yields a circular trajectory on a plane. The binormal correlation and the principal component analysis  reveal the butterfly, elliptical, and circular trajectories of the free end monomer. We present a phase diagram for different kinds of motion in the parameter regime of  compressive force and  torsional rigidity.
\end{abstract}
\maketitle

\section{Introduction}
Understanding the collective dynamics of active agents are much on focus in recent years. These active agents, such as mammalian heard, birds flock, colonies of microorganisms such as bacteria, collection of biological cells, and synthetic micro-swimmers, span a broad spectrum of length scales and timescales~\cite{Marchetti2013,elgeti2015physics,bechinger2016active,lauga2009hydrodynamics,cates2012diffusive,ramaswamy2010active,brennen1977fluid,palacci2010sedimentation,jiang2010active,bricard2013emergence,geyer2018sounds}. Though the elements of these active systems bear vast diversity in both individual character as well as in mutual interactions, one can capture much of the essential physics utilizing minimal models ~\cite{Romanczuk2012, bechinger2016active, lauga2007floppy, cates2013active, vicsek1995novel,toner1995long}. A subset of such minimal models is active filaments~\cite{ghosh2014dynamics, eisenstecken2016conformational, jiang2014motion, isele2015self, laskar2017filament, anand2018structure, isele2016dynamics,anand2019behavior}, which help to gain an understanding of a large class of problems involving elongated self-propelling elements~\cite{brennen1977fluid}. It is well known that a filament with finite flexibility, when constrained at one end, shows interesting oscillatory dynamics~\cite{chelakkot2014flagellar, Jayaraman2012, Laskar2013, de2017spontaneous, elgeti2013emergence, chakrabarti2019spontaneous,ling2018instability}. In studies which consider fluid mediated interactions, such oscillations are consequence of hydrodynamic instability~\cite{Laskar2013}. Interestingly, in `dry' systems where the fluid mediated interactions are not present, such oscillations arise as a result of an elastic instability, via `follower forces'~\cite{chelakkot2014flagellar,Fatehiboroujeni2018}.

A simple experimental realization of active filaments would be a linear chain of connected active particles, each of them propels along the local tangent of the chain~\cite{ghosh2014dynamics, eisenstecken2016conformational, jiang2014motion, isele2015self, laskar2017filament, anand2018structure, isele2016dynamics}.  The control parameter here is the self-propulsion speed of active particles, which causes compressive stress along the filament. These kind of active stresses have been observed in the system of microtubules and molecular motors, where a motor slides over the filament and causes motility. A collection of these  filaments and motors  on the surface shows various emergent phases and defects~\cite{doostmohammadi2018active,ndlec1997self,schaller2010polar,sanchez2012spontaneous}. If we impose translational and rotational restrictions on one end of the filament, beyond a threshold propulsion speed, the compressive stress causes the filament to buckle. The imposed restriction provides a coupling between the filament shape and the active compressive stresses, which results  oscillation of  the filament with a fixed frequency. Interestingly, the filament's oscillations caused by this mechanism have many qualitative similarities to the oscillations of eukaryotic flagellum~\cite{lindemann2010flagellar, lindemann2004testing, sleigh1968patterns,vilfan2019flagella}. The beating motion of eukaryotic flagella and cilia has significant functions in biology, in the context of locomotion of microorganisms, micro-scale fluid pumping in various organelles, development of embryos, etc~\cite{fulford1986muco, nonaka2002determination, sawamoto2006new,shields2010biometric}. Though eukaryotic flagella and cilia are highly complex in the structure, recent experiments suggest similar oscillations in much simpler in vitro systems~\cite{Sanchez2011}. More recently, experiments on the filaments made of synthetic active particles have shown flagella-like beating motion~\cite{Nishiguchi2018}. These recent advances in the development of artificial systems that mimic flagellar oscillations strengthen the possibility of experimental realization of micropumps based on the flagellar beating~\cite{dreyfus2005microscopic}. For designing an artificial system that mimics flagellar oscillations, the follower force mechanism is a natural candidate. It is therefore essential to systematically characterize the various dynamical regimes of models that provide oscillatory dynamics in the elastic filaments. 

Previous studies have well characterized the dynamical regimes of oscillations in two-dimensions. However, a systematic analysis of the dynamics as a function of various mechanical control parameters is still lacking. Further, to analyze a filament in 3D, one has to consider the torsional rigidity of the filament in addition to the bending rigidity. Here, we study the beating motion of a composite filament, realized by connecting three semi-flexible filaments in parallel using elastic spring potentials. The ribbon-like arrangement naturally provides torsional rigidity to the filament. In this article, we study the dynamics of an active ribbon-like filament in three dimensions. The filament model includes both bending and torsional rigidities, whose strength we can control using parameters of the rigidity potential. Our analysis reveals the influence of activity and ratio of bending and torsional rigidities of the filament on its  dynamics. 

The article is organized as follows: The simulation model of the clamped ribbon is discussed in section II. All the results are presented in section III. Results are concluded in the summary section IV. 

\section{Model}
\begin{figure}%[h]
	\includegraphics[width=\linewidth]{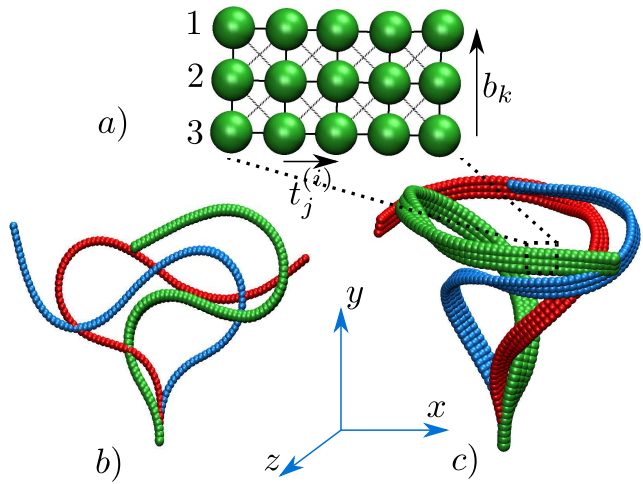}
	\caption{a) A schematic picture of the model of a ribbon  made from three prototype filaments with $k^{th}$ binormal vector (at $k^{th}$ monomer) and tangent vector $t^{(i)}_j$ on the $i^{th}$  protofilament ($i=3$) and at $j^{th}$ monomer.  Snapshots show different kind of motional phases: Beating (b) and Rotational phases (c), respectively. Different colors show the timeline of the events as: blue, red, and green at the end. Moreover 1,2, and 3 in (a) refers to  different protofilaments.  }
	\label{model}
\end{figure}

We consider a thin ribbon made of a parallel assembly of three stiff protofilaments (see Fig.~\ref{model}a). Each protofilament consists $N_{m}$ monomers with spatial position ${\bf r}_i$, connected by harmonic potential of the rest length $\ell_0$. To arrange the protofilaments into a ribbon, we also connect the monomers of adjacent protofilaments via quadratic elastic potentials. Each monomer is connected to its immediate neighbors (to its left and to its right) of the neighboring protofilaments via a harmonic potential of the rest length $\ell_0$. Additionally, we also implement diagonal connections between monomers of different protofilaments via harmonic potential of rest length $\sqrt{2}\ell_0$ (see Fig.~\ref{model}). This way, each monomer is connected to maximum eight other monomers in the ribbon. This arrangement ensures an equilibrium distance between the center-lines of a pair of adjacent protofilaments to be $l_0$, same as the equilibrium bond length in a protofilament. 

The bond potential energy is expressed as $U_{s}$, 
\begin{equation}
U_{s} = \frac{\kappa_{s}}{2}\sum_{i=1}^{3N_m}\left\{\sum_{j\in N_1}^{'}(|\textbf{R}_{i,j}|-\ell_{0})^2+
\sum_{k \in N_2}^{'} (|\textbf{R}_{i,k}|-\sqrt{2}\ell_0)^2 \right\}.
\label{Eq:spring}	
\end{equation}

Here, $\kappa_s$ is the spring constant and ${\bf R}_{i,j} = {\bf r}_i - {\bf r}_j$ is the bond vector between a pair of monomers. The summation in $j$ and $k$ are over nearest neighbors $N_1$ and next nearest neighbors $N_2$ of $i^{th}$ monomer. We impose two bending potentials to restrict curvatures in two directions. The first bending potential suppresses angular fluctuations between tangent vectors of each protofilament, such that for the ribbon,
\begin{equation}
U^{(1)}_{b} = \frac{\kappa_b}{2}\sum_{i=1}^{3}\sum_{j=1}^{N_m-2} \left(\textbf{t}_{j+1}^{(i)}-\textbf{t}_j^{(i)}\right)^2,
\label{bending_1}
\end{equation}
where $\textbf{t}_j^{(i)} = \textbf{R}_{j,j+1}^{(i)}$ is the $j^{th}$ tangent vector of the $i^{\text{th}}$ protofilament. We impose another  bending potential between three monomers of different protofilaments, which have the same height from the clamped end at equilibrium. The form of potential energy is  
\begin{equation}
U^{(2)}_{b} =  \frac{\kappa_b}{2}\sum_{j=1}^{N_m} \left(\textbf{d}_{j}^{(2,1)}-\textbf{d}_{j}^{(3,2)}\right)^2,
\end{equation}
where $\textbf{d}^{(i+1,i)}_j = \left(\textbf{r}^{(i+1)}_j - \textbf{r}^{(i)}_j \right)$,  is the vector connecting $j^{th}$ monomers of the protofilements $i$ and $i+1$. Apart from the bending, we also impose a torsional potential energy $U_{t}$ in a simplified manner. This is treated as the bending potential, which penalize the angular displacement between vectors (${\bf d}_i$) on the effective bi-normal direction of the ribbon, such that ${\bf b}_i = \textbf{d}_{i}^{(2,1)}+\textbf{d}_{i}^{(3,2)}= \textbf{r}^{(3)}_i - \textbf{r}^{(1)}_i$, therefore the torsion energy is given as
\begin{equation}
U_{t} = \frac{\kappa_{t}}{2} \sum_{i=1}^{N_m-1}({\bf b}_{i+1} - {\bf b}_i )^2 .
\label{Eq:torsion}
\end{equation}
To impose self-avoidance among protofilaments, we implement excluded volume interaction between monomers via WCA potential such that the pair of monomers ($i,j$) separated by a distance $R_{i,j}$, $U_{ij} = 4\epsilon [(\sigma/R_{i,j})^{12} - (\sigma/R_{i,j})^6 + 1/4]$ if $R_{i,j} < 2^{1/6}\sigma$ and zero otherwise. The activity causes a compressive stress on the ribbon by applying a force on the monomers of the centre protofilament, of the form as $\textbf{F}^{i}_{a}=-f_{a} {\bf t}^{(2)}_i$, where $f_{a}$ is the strength of active force~\cite{anand2018structure,isele2015self,chelakkot2014flagellar}.

The equation of motion for a  monomer of the filament in the overdamped limit is,
\begin{equation}
\gamma \frac{d {\textbf{r}_i}}{d t} = - \nabla_{i} U + \textbf{F}_{r}^{i} + \textbf{F}_{a}^{i},
\label{eq:langevin}
\end{equation}
here $\gamma$ is the friction coefficient, $\textbf{F}_{r}^{i}$ is  white noise  with zero mean, and $U$ is total potential energy of the filament, given as $U = U_{s} + U^1_{b} +U^2_{b} +U_{t} + U_{LJ}$.  The viscous drag and the noise are related through the fluctuation-dissipation relation, 
%\begin{eqnarray}
$\langle \textbf{F}_{r}^{i}(t)\cdot \textbf{F}_{r}^{j}(t')
\rangle  =  6k_{B}T\gamma \delta_{ij}\delta(t-t')$.
%\label{eqn-7}.
%\end{eqnarray}
We use Euler integration technique to solve the equation of motion with integration step size to be in the range of $h_m=10^{-3}\tau$ to $10^{-5}\tau$. We arrange the protofilaments in the $z$ direction with $y$ to be the vertical direction. Thus, the $y-z$ plane is fixed as plane of the ribbon, whereas $x-y$ plane is the beating plane,  which will be discussed in the latter sections. First two monomers of each protofilaments at the basal end of the ribbon (see Fig.~\ref{model}-b) are clamped in the vertical direction.

The physical parameters such as, frequency $\omega$, various energies, forces, time, bending  and  spring constants, and various lengths are presented  in units of the bond length $\ell_0$, diffusion coefficient of a monomer $D_m=k_BT/\gamma$,  and thermal energy $k_{B}T$. Each protofilament consists of $N_m=100$ monomers, thus total $N=300$ monomers for the ribbon. The parameters are chosen as, $\sigma=\ell_0$, $\epsilon/k_{B}T=1$, and time is in units of $t_s = \ell_0^2/D_m=1$. The spring and  bending parameters are taken to be $k_{s}=1000 k_BT/\ell_0^2$, and $\kappa_{b}=1000k_BT/\ell_0^2$, respectively. The torsion parameter $\kappa_{t}$ is varied in the range from $0$ to $\kappa_{b}$. The strength of torsion parameter is expressed in terms of  ratio of torsion to bending rigidities given as $\rho= \kappa_{t}/\kappa_{b}$. Thus, $\rho \ll 1$ corresponds to small torsional rigidity and $\rho \approx 1$ corresponds to large torsional rigidity. Even when $\rho =0$, the model has a non-zero torsional resistance due to the elastic force between the protofilaments. However in this specific model, $\rho$ acts as a useful control parameter to tune the torsional rigidity, which can be directly quantified by measuring the  bi-normal persistence length ($L_b$) from the relaxation of bi-normal vector of the ribbon, $C_b(s)=\langle{\textbf{b}}_{s+s_0 }\cdot {\textbf{b}}_{s_0}\rangle\sim \exp(-s/L_b)$  ~\cite{giomi2010statistical}. Here, binormal vector $b_i$ at $i^{th}$ monomer is presented as $b_s \equiv b_i$ with $s\simeq i\ell_0$  where arc length $s$ varies  in dimensionless unit from $0$ to $100$,  and  $s_0$ is always taken at the basal end, i.e.,  $s_0=0$. We obtain that $L_b$ is linear function of $\rho$ (see Fig.~SI-1a).
The strength of active force is represented in terms of a dimensionless (P{\'e}clet) number $Pe=f_al_0/k_BT$, which we vary from $0$ to $20$ in our simulations. Each physical quantity is averaged over 40 independent realizations. The hydrodynamic interactions among the monomers of the protofilaments  are neglected here for the simplicity of calculation.

\begin{figure}[t]
	\includegraphics[width=\linewidth]{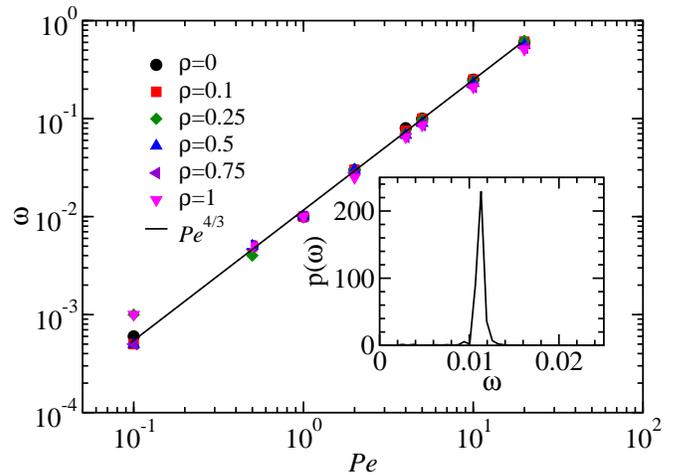}
	\caption{Dependence of frequency on the active force. The solid line illustrates the power law behavior of the oscillation frequency.  Inset shows power-spectrum of the projection of end monomer on the clamped plane for $\rho=0.25$ and $f_{a}=1$.}
	\label{Fig:time-series}
\end{figure}

\section{Results}

A straight vertically clamped ribbon goes through buckling transition when the load or compressive force  density $f_a$ is larger than a critical value. It is well known that for an active filament in 2D, the buckling transition leads to an oscillatory motion since the local active force follows the local tangent~\cite{chelakkot2014flagellar,sarkar2017spontaneous,yang2018beating,saggiorato2017human}. However, in the case of a ribbon, the torsional rigidity contributes an additional elastic component which acts against induced deformations. This factor modifies the periodic oscillations observed in filaments without torsional rigidity and leads to new dynamical states. We systematically study the filament's dynamics as a function of $Pe$ by varying the torsional contribution to total elasticity by changing the parameter $\rho$.  For a sufficiently large $Pe$, we observe in-plane periodic oscillations (Fig~\ref{model}-b) in the limit of large torsional rigidity $\rho \rightarrow 1$, whereas for negligible torsional rigidity, we observe out-of-plane, circular oscillations (Fig~\ref{model}-c).  Further analysis reveals that the free end of ribbon exhibits butterfly and elliptical trajectories for the intermediate values of $\rho$. We also quantify these phases by calculating binormal correlations and the distribution of the angle made by the clamped-to-end vector to the vertical axis.

\subsection{Periodic motion of filament}
% Beyond a threshold value of $Pe$, the compressive force on the filament is larger than the critical buckling force. At these values of $Pe$, the ribbon starts to oscillate periodically about the vertical axis with a definite frequency. As has been reported previously in the case of beating  of filaments~\cite{chelakkot2014flagellar,sarkar2017spontaneous,yang2018beating,saggiorato2017human}, 
To quantify  oscillation of ribbon for large $Pe$ and to calculate the frequency, we record the time-series data of the deflection of the end in the $x$ direction. For planar oscillations (when $\rho \rightarrow 1$), the $x$ deflection provides the oscillation amplitude of the end segment, whereas in the case of non-planar oscillations (when $\rho \ll 1$) this quantity is useful to analyse the oscillation cycles. In Fourier space, the peak of the power-spectrum of this time-series provides the frequency and ascertains the oscillatory behavior of the filament (Fig.~\ref{Fig:time-series}). The peak identifies the oscillation frequency ($\omega$) for the corresponding time-series data. We estimate $\omega$ for different strengths of $Pe$ as well as for different torsion parameters. We find that $\omega$ grows with $Pe$ and follows the scaling relation $\omega \sim Pe^{4/3}$ as reported earlier~\cite{chelakkot2014flagellar,sarkar2017spontaneous}. Interestingly, $\omega$ shows  a weak dependence on the torsion rigidity parameter $\rho=\kappa_{t}/\kappa_b$. As reflected in Fig.~\ref{Fig:time-series}, the values of oscillation frequencies are nearly same for a given $Pe$ for all values of $\rho$.

Our analysis reveals that the scaling behavior  $\omega \sim Pe^{4/3}$ is retained even for non-planar oscillations when $\rho \ll 1$. 
The same scaling behavior is obtained in planar beating of a filament in 2D, also due to follower force mechanism~\cite{chelakkot2014flagellar}. In the case of 2D filament, the scaling behavior can be derived from a balance of energy dissipation due to viscous friction and the energy input from active forces over a characteristic length-scale $\lambda$ and a characteristic time $\omega^{-1}$. Here the time-scale $\omega^{-1}$ is given by the period of oscillation. The length-scale $\lambda$ is the bending length-scale which scales as $(\kappa/Pe)^{1/3}$. Therefore, the scaling $\omega \sim Pe^{4/3}$ indicates that the oscillations in $x$ deflection is same as bending oscillations observed in 2D filaments. To check the presence of additional active oscillations other than bending at $\rho \ll 1$, we compute the time evolution of (i) the azimuthal angle of the filament end during the rotational motion and ii) the local torsional parameter $\chi^t_i$ and calculate their oscillation frequency (SI). However, all these frequencies coincide with the bending frequency. Our analysis confirms that the bending oscillation are the dominant mode of oscillation which controls all other types of oscillations in the system.

%{ \color{red} This scaling behavior indicates that the bending length-scale $\lambda$ of the ribbon is related to the active tangential stress via $\lambda \propto Pe^{-1/3}$  as in the case of planar filaments~\cite{chelakkot2014flagellar}. Then, the oscillation frequency is entirely determined by the balance between the viscous energy dissipation,  and the energy input due to active stress, over the length $\lambda$ in time $\omega^{-1}$ which is independent of the trajectory of the ribbon, whether it is planar or circular.}

\begin{figure}[t]
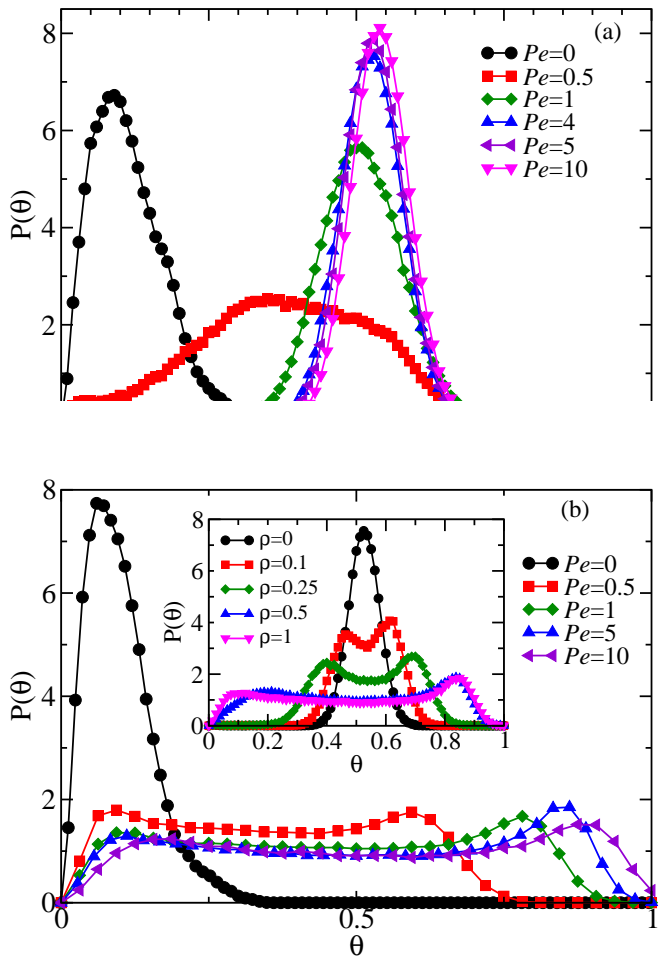

	\includegraphics[width=\linewidth]{angle_end_R0}
	\includegraphics[width=\linewidth]{angle_end_r100}
	\caption{Distribution of angles of clamped-to-free end vector from the clamped axis for a) $\rho=0$ and b) $\rho=1$. Inset of figure b) displays distribution of $\theta$ at $Pe=4$ for various $\rho$.}
	\label{Fig:dist R0}
\end{figure}

\subsection{Trajectory of Periodic motion}
The power spectrum of clamped-to-free end distance confirms the periodic nature of the motion and the scaling relation of $\omega$ with $Pe$. We further extend our analysis to distinguish in-plane and out-of-plane motion in detail by quantifying the trajectory of oscillations. We first calculate the angle between clamped-to-free end vector of the ribbon and the vertical axis (y-axis) such that, $\theta = \cos^{-1}(\hat{\bf e}.\hat{\bf {e}}_y)$, where $\hat{\bf e}$ is the unit vector along clamped-to-free end of the ribbon. For a perfectly straight ribbon along the vertical axis, $\theta$ is zero. However in equilibrium, thermal fluctuations cause weak bending about the vertical axis thus leading to non-zero, albeit small value for average $\theta$. Thus, for $Pe=0$, the distribution of $\theta$ ($\text{P}(\theta)$) displays a maximum around $\theta \simeq 0.1$ (in radian) as shown in (Fig~\ref{Fig:dist R0}). In the case of $Pe > 1$, the active compressive force causes the filament to bend more, leading to a larger $\theta$ and a significantly different P($\theta$). When $\rho=0$, we observe a gradual shift in the peak position of $P(\theta)$ with $Pe$, which saturates around $\theta \simeq 0.6$ at large values of $Pe$ (Fig.~\ref{Fig:dist R0}-a). A single peak in $P(\theta)$ and a well defined $\omega$, as evident from the power-spectral density, confirms that the end-to-end vector of the ribbon follows a cone with an average angle $\theta$ in the absence of torsional rigidity ($\rho=0$). This is also visible from the simulation video (SI-MOVIE-1). For high torsional rigidity {\it i.e.} $\rho\simeq1$, the distribution $\text{P}(\theta)$ is qualitatively different for $Pe >1$ in comparison with ribbon with $\rho =0$ (see Fig.~\ref{Fig:dist R0}). In this case, $\text{P}(\theta)$ is nearly uniformly distributed in a wide range of $\theta$, and this range broadens with an increase in $Pe$. The broad and nearly uniform distribution of $\theta$ indicates a planar motion of the clamped-to-free end vector and suggests a planar beating motion of the ribbon as observed in the case of 2D filaments (see SI-MOVIE-2). To unveil the role of $\rho$, we plot $P(\theta)$ for various $\rho$'s at a fixed $Pe$ in the inset of Fig.~\ref{Fig:dist R0}. We observe that the peak in distribution  broadens with $\rho$ in the range of $\rho=0$ to $\rho=1$. A  peak at $\rho=0$ becomes almost flat for  $\rho >0.25$, which suggests the transition in its motion.

Two asymptotic limits in terms of torsional rigidity are $\rho=0$ and $\rho=1$. Here, we observe an out-of-plane, rotational motion ($\rho=0$) where the end of the ribbon follows a circular trajectory in the $x-z$ plane or an in-plane ($\rho=1$), beating motion of the ribbon in the $x-y$ plane. However when $0<\rho<1$, we observe a series of complex dynamical phase of the ribbon. We analyse these phases by characterizing the trajectory of the end-segment with the help of principal component analysis~\cite{PCA_jollife,werner2014shape} (PCA). Here the coordinates of the end monomer of the middle filament are transformed according to PCA~\cite{PCA_jollife, stephens2008dimensionality,ma2014active}. We find two of the principal coordinates $X_1$ and $X_2$, which corresponds to the eigenvectors of the largest eigenvalues of the covariance matrix. The magnitude of these two eigenvalues dominates over the others. These two eigenvalues contribute nearly $90\%$ in the sum of squares of all eigenvalues.  In such a case, dynamics can be expressed in terms of these two eigenmodes.  Thus, PCA can help us to identify a  predominant plane of motion of the free end.  

\begin{figure*}
	\includegraphics[width=0.32\linewidth]{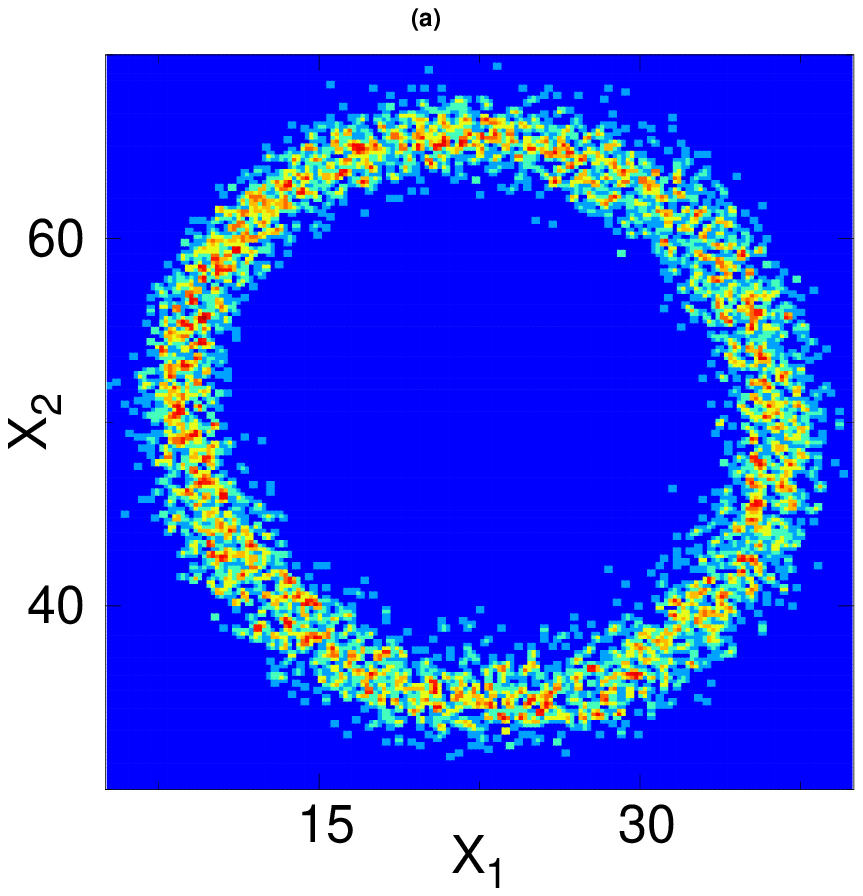}
	\includegraphics[width=0.32\linewidth]{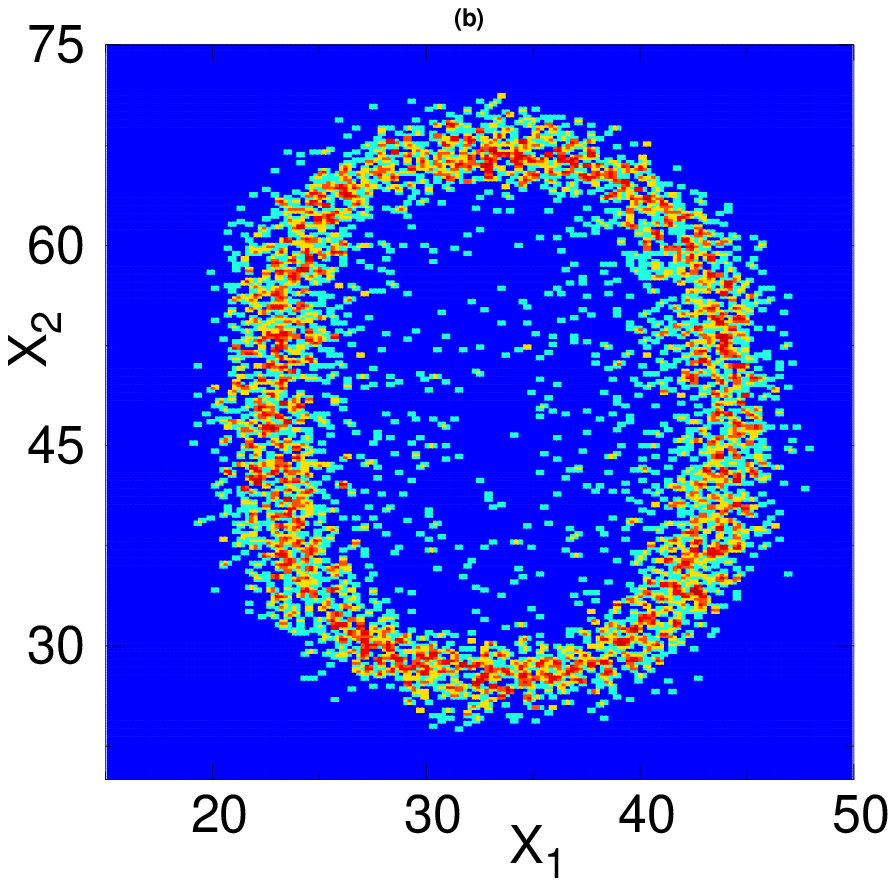}
	\includegraphics[width=0.32\linewidth]{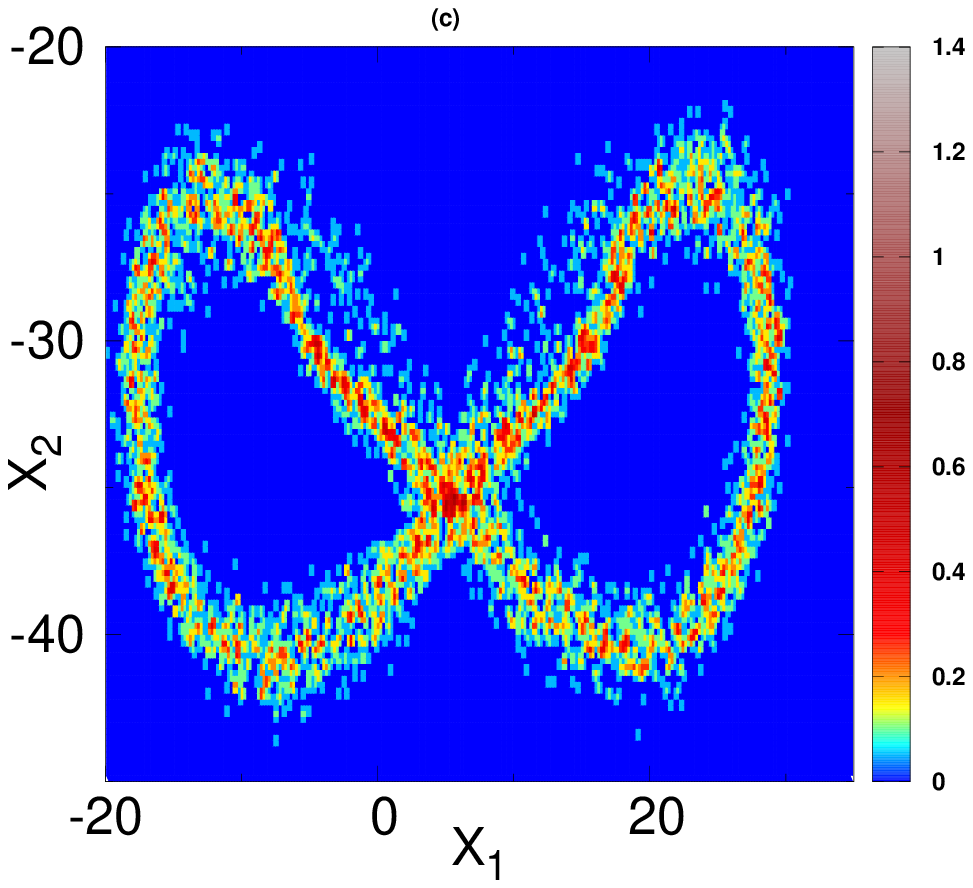}
	\caption{The distribution of largest two eigen-modes  of the end monomer in the PCA transformed space at $Pe=10$,  for $\rho=0$ (a), $\rho=0.25$ (b) and $\rho=1$ (c) from left to right.}
	\label{Fig:dist_pca}
\end{figure*}

\begin{figure}
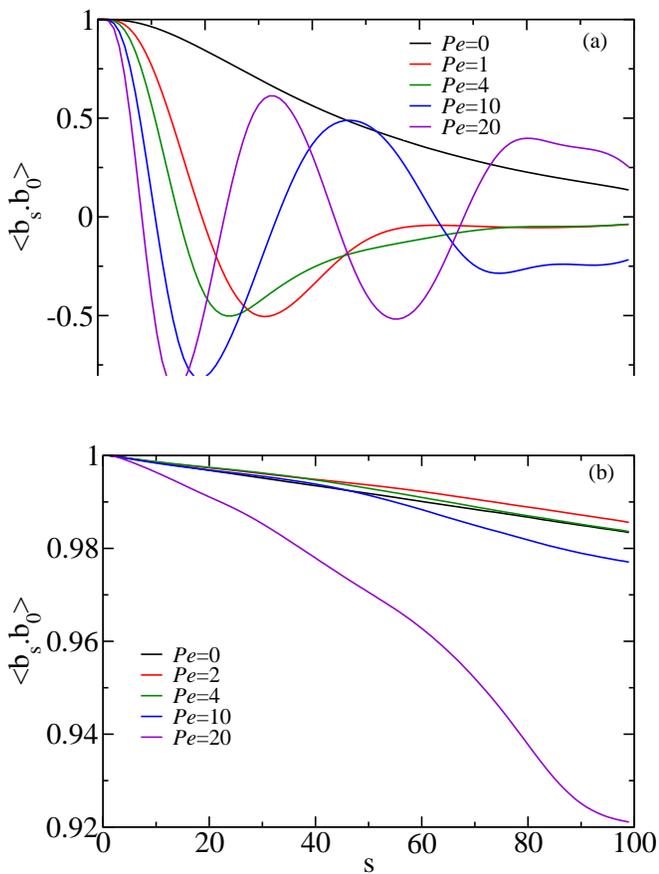

	\includegraphics[width=\linewidth]{binorm_R0}
	\includegraphics[width=\linewidth]{binorm_R100}
	\caption{The correlation of binormal vector $C_b(s)$ for the torsional parameters $\rho=0$ (a) and $\rho=1$ (b). Here, $s=0$ indicates the clamped end of the filament.}
	\label{Fig:binormal}
\end{figure}

Figure~\ref{Fig:dist_pca} displays the distribution of transformed coordinates of end monomer in $X_1$ and $X_2$ space for various range of $\rho$ at fixed $Pe=10$. Thus, it shows a continuous change in  trajectory from a circular to an elliptical shape with an increase in $\rho$, and then to a butterfly shape at large $\rho$ ( see Fig.~\ref{Fig:dist_pca}~a,b, and c). The transition from the circular motion to the butterfly motion is mediated by the distortion of a circular to  an elliptical  shape with torsion rigidity in the range of $\rho=0$ to  $0.25$. 
The change in the trajectory is linked with the kind of dynamical phases of the ribbon. The circular trajectory illustrates the rotational motion, whereas the butterfly shape assists our claim of a planar motion of the filament in the range of higher torsional rigidity. In the intermediate regime of $\rho$, transition from beating to rotational motion leads to a large scale transformation of the trajectory.

The shapes in Fig.~\ref{Fig:dist_pca} can be understood from the oscillation frequencies of PCA components and phase difference between their periodic motions as very much similar to Lissajous figures. 
Here two dominant principal components  can be assumed as  $X_{1}=a \sin(\omega_{1}t+\phi_p)$ and $X_{2}=b \sin(\omega_{2}t)$. In case of $\omega_{1}/\omega_{2}=1$ and $\phi_p=\pi/2$, the phase space of $X_{1}$ and $X_{2}$ becomes a circle/ellipse, which we see in Fig.~\ref{Fig:dist_pca}-a and b. Similarly,  for the parameters $\phi_p=0$ and $\omega_{1}/\omega_{2}=1/2$ it traces the butterfly shape on $X_1$ and $X_2$ plane as shown in  Fig.~\ref{Fig:dist_pca}-c in the beating phase.
%In the parameter regime where the filament rotates around its equilibrium configuration, both the frequencies are same. But the frequency of $X_{2}$ becomes twice of the frequency of $X_{1}$ in the beating phase. 

\subsection{Binormal Correlation}

Our analysis of the ribbon trajectory shows that the out-of-plane dynamics of the ribbon is suppressed in the limit of $\rho \rightarrow 1$. For a given $\rho$, the effective torsional rigidity of the ribbon in equilibrium can be estimated by measuring the binormal vector correlation $C_b(s)=\langle{\textbf{b}}_{s+s_0} \cdot {\textbf{b}}_{s_0} \rangle$. The correlation of binormal vector decays exponentially  along the contour for $Pe=0$, i.e., $C_b(s) \sim \exp(-s/L_b)$, where $L_b$ is the effective binormal persistence length~\cite{giomi2010statistical,golestanian2000statistical,liverpool1998statistical}.

Now, we quantify the effect of active, compressive force on the binormal-binormal correlation of a clamped ribbon, by calculating $C_b(s)$ for various values of $Pe$ and $\rho$. We find that $C_b(s)$ is qualitatively different for non-zero $Pe$ at $\rho=0$ in comparison to the passive ribbon, as it shows oscillatory behavior as a function of $s$ (see Fig.~\ref{Fig:binormal}-a). The oscillatory behavior in $C_b$ persists even for non-zero values of $\rho$, until $\rho \lesssim 0.25$. Spatial oscillations in $C_b$ indicates twist deformations~\cite{golestanian2000statistical,giomi2010statistical}. As the appearance of these oscillations coincides with the out-of-plane movement of the ribbon, we deduce that the out-of-plane movement of the ribbon is accompanied by significant twisting of the ribbon. Therefore, one can use $C_b$ as another indicator for the out-of-plane dynamics of the ribbon. For $\rho \simeq 1$, we find $C_b$ decays exponentially with $s$ even at large $Pe$, indicating negligible twisting of the ribbon. The ribbon displays in-plane oscillations in this regime as Fig.~\ref{Fig:binormal}-b illustrates.  

\subsection{Dynamical Phases}

\begin{figure}[t]
	\includegraphics[width=0.5\linewidth,height=5cm]{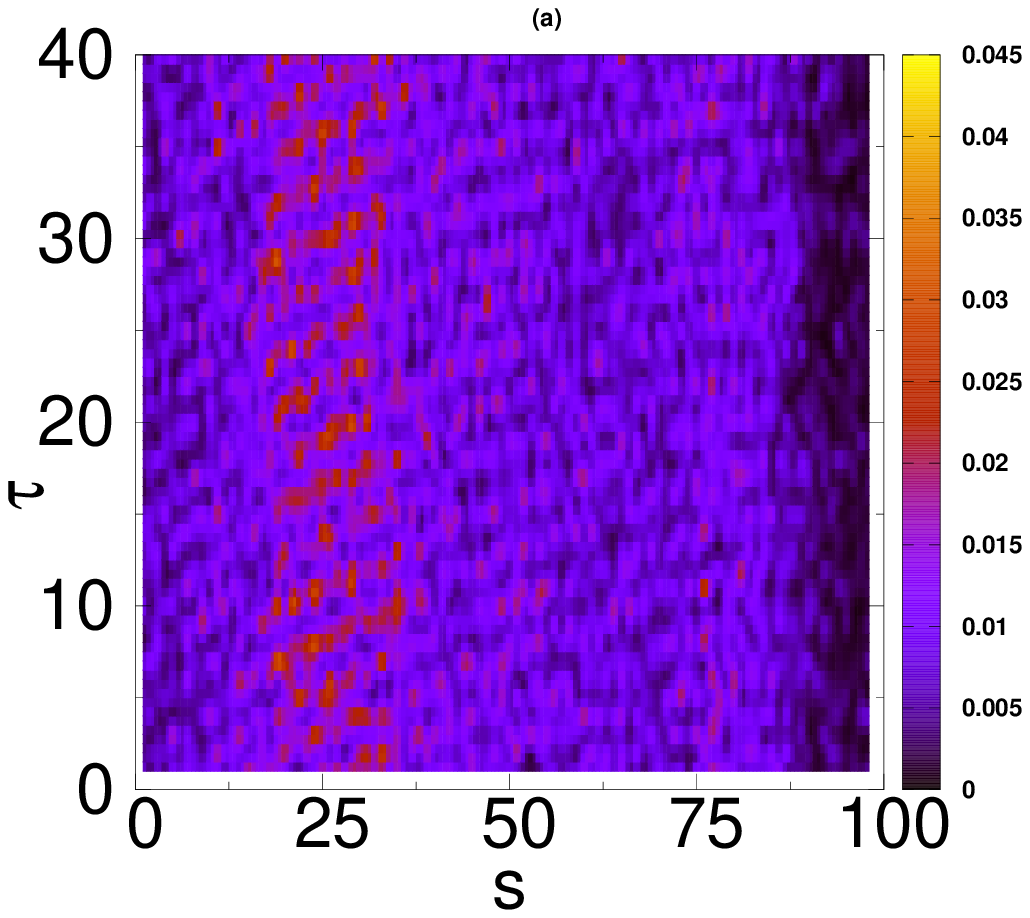}
	\includegraphics[width=0.48\linewidth,height=5cm]{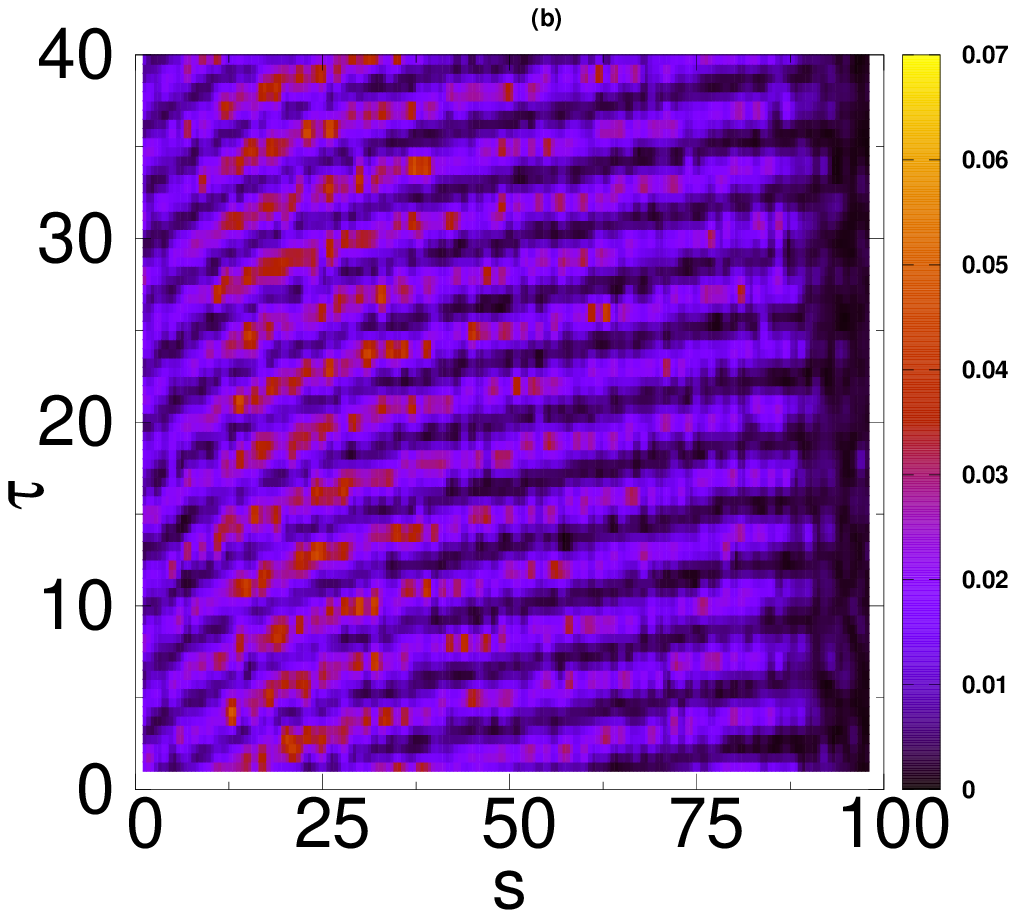} \\
	\includegraphics[width=0.5\linewidth,height=5cm]{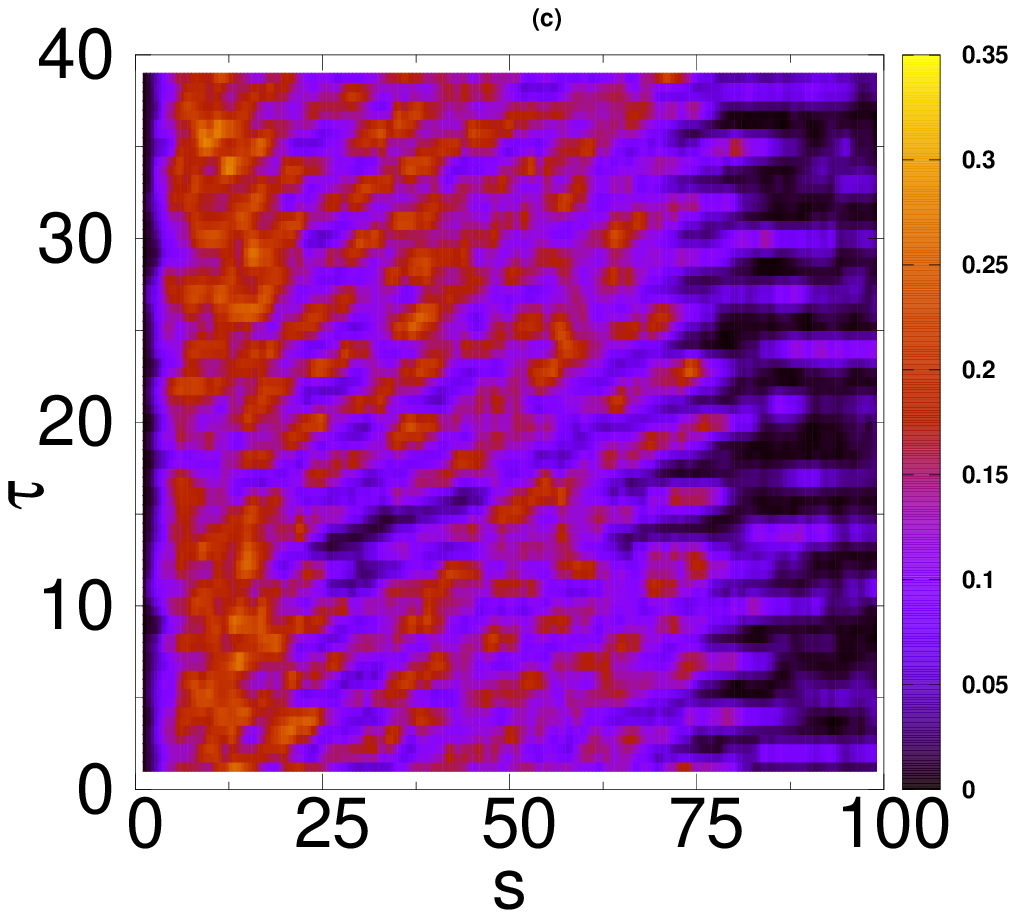}
	\includegraphics[width=0.48\linewidth,height=5cm]{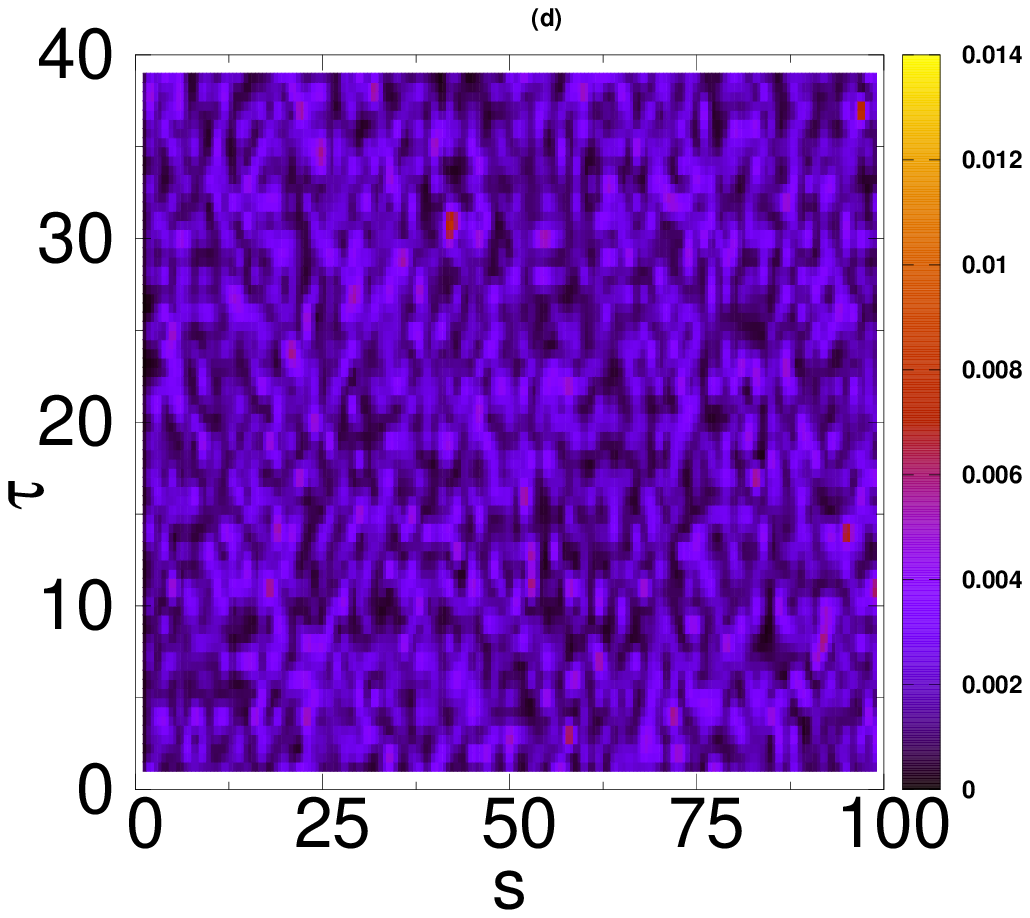}
	\caption{ Kymograph for the bending energy for ${\rho}=0$ (a) and ${\rho}=1$ (b), respectively at $Pe=10$. Kymograph for the torsion parameter for $\rho=0$ (c), and $\rho=1$ (d), respectively at $Pe=10$. Time ($\tau$) is along  y-axis  and  x-axis is the arc length $s$ in all plots.}
	\label{Fig:bending}
\end{figure}

\begin{figure}[t]
	\includegraphics[width=\linewidth]{twist_energy}
	\includegraphics[width=\linewidth]{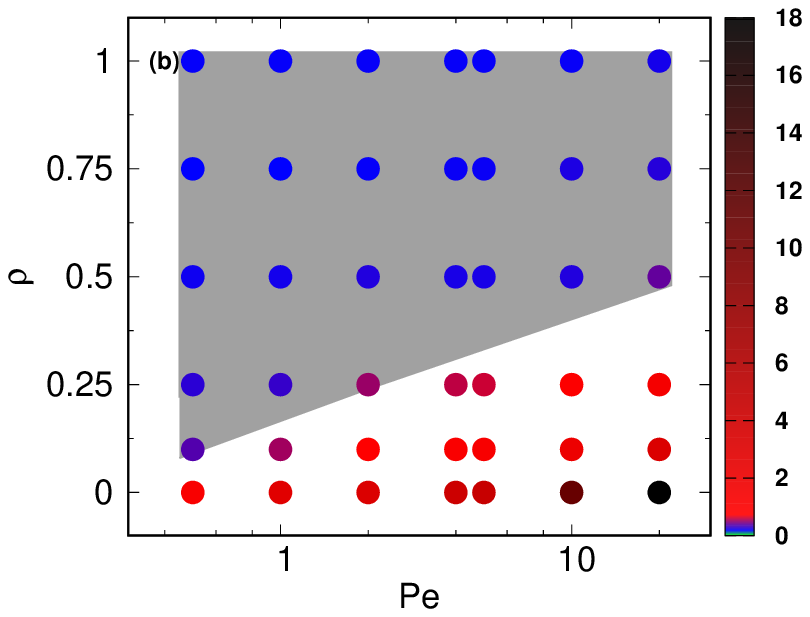}
	\caption{a) The torsion parameter $\chi_{t}$ as a function of $Pe$ for various $\rho$. b) Phase-diagram showing different dynamical phases in the parameter space. Grey shaded area shows the parameter regime for the beating motion, where as rest of the area shows rotational regime. Color code for circles is based on the torsion parameter $\chi_{t}$. Stationary states at low $Pe$ are not included.}
	\label{Fig:phase}
\end{figure}

In the previous section, we have identified the nature of the dynamics of the ribbon through its terminus as a function $Pe$ and $\rho$. Here, we summarize dynamics of the clamped active ribbon in the parameter space of $\rho$ and $Pe$. The in-plane-motion is accompanied by a significant planar bending of the ribbon, whereas the out-of-plane dynamics causes twisting of the ribbon.  We quantify these states by calculating two local geometric quantities, the bending parameter ($\chi^{(b)}_{i}$) and the twisting parameter ($\chi_i^{(t)}$). The local bending parameter is given as $\chi_i^{(b)} = \sum_{j=1}^3 (\textbf{t}_{j}^{i+1}-\textbf{t}_{j}^i)^2$, where $\textbf{t}_j^{i}$ is the $i^{th}$ tangent vector of the $j^{th}$ protofilament. The local torsional parameter $\chi_i^{(t)} = ({\bf b}_{i+1} - {\bf b}_i)^2$, where ${\bf b}_i$ is the local binormal vector as shown in the Fig.\ref{model}-a. We plot the $\chi_i^{(b)}$ and $\chi_i^{(t)}$ in the form of a kymograph, which provides the spatio-temporal variation of the respective quantities in Fig.\ref{Fig:bending}. For $Pe=10$, the bending kymograph indicates propagation of the bending waves from the basal end to the free end of the filament for both $\rho=0$ and $\rho=1$, indicating that the local bending energy of the ribbon oscillates periodically for both in-plane and out-of-plane oscillations. However, the torsional kymograph shows no pattern in the case of planar oscillations when $\rho=1$, whereas it indicates propagation of periodic `twisting waves' in the case of out-of-plane oscillations when $\rho=0$.

The spatio-temporal pattern formed by both torsional and bending parameters confirms the presence of out-of-plane motion for $\rho \lesssim 0.25$ and beating motion for large values of $\rho$. In order to quantitatively distinguish between in-plane and out-of-plane dynamics of the ribbon,  we define a global torsional parameter $\chi_{t} = \sum_i^{N_m} \chi^{(t)}_i $.	It is evident from Fig.~\ref{Fig:phase}-a that $\chi_{t}$ remains close to zero in a planar motion, but becomes a large number for out-of-plane movements. We find that $\chi_{t}$ increases linearly for small $\rho=0,0.1 $, and $0.25$ (see Fig.~\ref{Fig:phase}-a). In the limit of $\rho>0.25$, $\chi_t$ is almost constant with $Pe$ and exhibits negligible change. Thus, $\chi_{t}$ serves as a good indicator for the out of plane motion of the ribbon.

We use the magnitude of geometric parameter $\chi_{t}$ as a parameter to quantify the out-of-plane movement in the parameter space of $Pe$ and $\rho$. Figure~\ref{Fig:phase}-b displays a phase diagram for dynamical phases with  help of color map, which  shows variation in $\chi_{t}$ in different phases. In the diagram, we only indicate those points which provide oscillatory dynamics, i.e., only the values of $Pe$ larger than the critical value. The grey shaded area corresponds to in-plane beating motion, whereas the unshaded region corresponds to the rotational phase. From the phase curve, it shows the large global torsional  parameter  in the rotational phase (red) and small in the beating phase (blue). 

\subsection{Role of anisotropic friction}

Hydrodynamic interactions are ignored in our simulation, however it plays an important role for the active matter and self-propelled systems~\cite{pooley2007hydrodynamic,yang2008cooperation,brumley2014flagellar}. In case of the rod-like filament, friction is anisotropic in the presence of hydrodynamic interactions. Our approach assumes  friction to be isotropic thus diffusion too. To verify  the universality of our results, we incorporate  anisotropic friction to be in the same framework of our simulations, in the absence of hydrodynamic interactions. Thus, the friction parallel and perpendicular to the bond-vectors are taken to be $\gamma_{\perp}=2\gamma_{\parallel}$, similar to a filament in solvent~\cite{tirado1984comparison,doi1988theory}. This effectively modifies diffusivity of the filament in both directions, and mimics the role of hydrodynamics in a very averaged manner. We present here results for  $\rho$ at $0,0.25$, and $1$, and vary P{\'e}clet number to study the behavior of the filament with new protocol.

\begin{figure}%[h]
	\includegraphics[width=\linewidth]{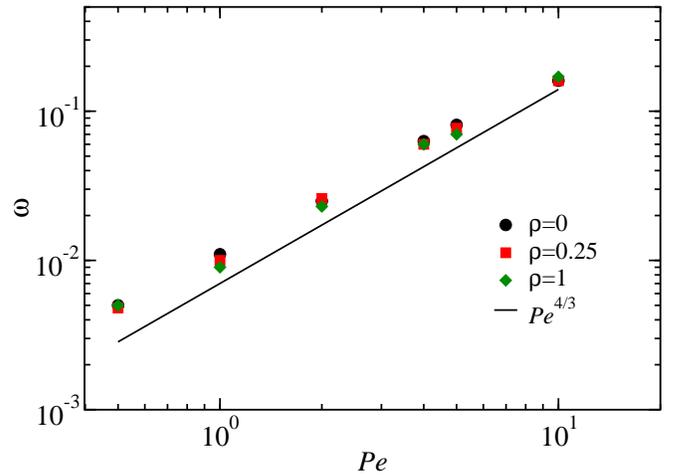}
	\caption{The frequency of periodic motion of the filament with  active force for the anisotropic friction 
		for various strength of torsion parameter $\rho$. Solid line shows the 
		power law behavior of the curves as, $\omega \sim Pe^{4/3}$.}
	\label{Fig:anisotropic}
\end{figure}

We present the oscillation frequency of the filament  in Fig.~\ref{Fig:anisotropic} under assumption of anisotropic friction.  It displays that oscillation frequency $\omega$ varies with power law as $\omega\sim Pe^{4/3}$ with exponent similar to the isotropic friction case, for $\rho=0,0.25$ and $1.0$. Although values of the oscillation frequencies are different than the isotropic case, however the characteristic features  of the results are same. The beating phase and rotational motion is also observed here, thus the results presented in this article are consistent with the anisotropic friction of the filament.

\section{Summary and conclusions}

In this article, we have presented a systematic study of the dynamics of an active ribbon, clamped at one end which oscillates due to the follower force mechanism. 
We have identified the mechanical regime in which the modelled ribbon mimics the flagellar motion.  We have also analyzed both in-plane and out-of-plane oscillatory motions of the ribbon.  The torsion parameter acts as a control parameter in the model, which dictates the transition from in-plane to out-of-plane movement.

We have characterized the periodic oscillations of the ribbon by three different measurements. First, we have calculated the oscillation frequency, which follows the scaling relation $\omega \sim Pe^{4/3}$ for all values of the torsional parameter $\rho$. We observed only a weak dependence of $\omega$ on $\rho$. The qualitative difference in the oscillations is manifested in the distributions of angle between clamped-to-free end vector and vertical axis. While the planar beating motion is linked to a broad distribution of angle $\theta$, the rotational motion leads to a peak about an angle $\theta > 0$. The width of the distribution increases with  $Pe$. Third, we have visualized the transition from in-plane to  out-of-plane motion by analyzing the trajectories of the free-end segment of the ribbon on the PCA transformed planes. By this, we have shown that the trajectory of the free end of the ribbon exhibits intricate dynamical patterns, including butterfly, elliptical, and circular trajectories.

Finally, we have identified the regions in the parameter space defined by $Pe$ and $\rho$, where different types of oscillations take place. For this purpose, we have estimated the binormal correlation $C_b(s)$ and a geometric torsional parameter $\chi_{t}$. Oscillations in $C_b(s)$ identifies the rotation of the filament.  Our analysis reveals that the planar beating and the rotational motion depend on torsional rigidity of the ribbon (via $\rho$) as well as the compressive force (via $Pe$). For small values of $\rho$, filament shows rotational motion  for  all values of $Pe$, whereas at large $\rho$ the ribbon exhibits beating motion. At the intermediate values of $\rho$, the oscillatory behavior depends crucially on the magnitude of $Pe$. We have also shown that torsional energy shows periodic oscillations as a consequence of the rotation of the filament, which disappears in beating phase, whereas planar bending energy always exhibits periodic behavior. 	

Our study provides an insight into how the oscillatory dynamics of a beating filament changes qualitatively by altering the strength of internal elastic elements. This study helps to design the synthesis of artificial flagella/cilia to be used in micro-scale structures. Although the internal driving mechanism of natural flagella/cilia are fundamentally different, our study hints at the importance of the arrangement of the structural elements in determining their planar beating dynamics. We also show that by tuning the torsional rigidity of a beating filament, one can qualitatively alter the oscillatory pattern of a filament. An eukaryotic flagellum beats due to the active stress generated on the axoneme by molecular motors ~\cite{lindemann2010flagellar,lindemann2004testing,brokaw1971bend,cibert2004geometry}. Simulation models including explicit coarse-grained motors acting on  the assembly of filaments may provide more insights into dynamical states of the axoneme. It's worth to  consider such systems in the future studies in a more intricate  manner. 

\section{Acknowledgements}
The computation work is carried out at HPC facility in IISER Bhopal. SPS acknowledges DST SERB Grant No. YSS/2015/000230 for financial support. RC acknowledges DST SERB for financial support via Ramanujan fellowship No. SB/S2/RJN-051/2015. 

%%%END OF MAIN TEXT%%%
%\balance

%\bibliographystyle{apsrev}
%\bibliography{citation}
%----------------------------------CITATIONS-------------------------------------------------------------
%--------------------------------------------------------------------------------------------------------

%--------------------------------------------------------------------------------------------------------
%--------------------------------------------------------------------------------------------------------

\pagebreak
\renewcommand{\thefigure}{SI-\arabic{figure}}
\setcounter{figure}{0}

\section*{Supplementary Text}
\begin{figure}[h]
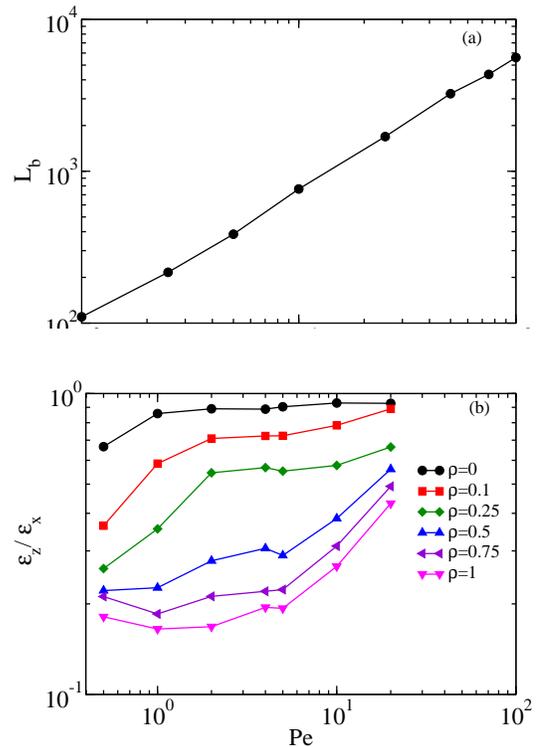

	\includegraphics[width=0.8\columnwidth]{pers_ratios}
	\includegraphics[width=0.8\columnwidth]{strain_ratio}
	\caption{a) Binormal persistence length for various $\rho=\kappa_{t}/\kappa_{b}$ ratio. b) Ratio of amplitudes in the direction of beating plane and perpendicular to the beating plane.}
	\label{Fig:persist}
\end{figure}

The binormal persistence length ($L_{b}$) of the ribbon for various torsional strength $\rho$ is displayed in the Fig.~\ref{Fig:persist}. It is estimated from the binormal-binormal correlation given as,  $C_b(s)=\langle{\textbf{b}}_{s+s_0} \cdot {\textbf{b}}_{s_0} \rangle\sim \exp(-s/L_b)$. Interestingly, the persistence length varies linearly with $\rho$. The increase in the  persistence length suggests the strengthening of torsional rigidity of the ribbon with $\rho$.

The ratio of mean of amplitudes of the periodic motion on the x-z plane defined as $\epsilon_z/\epsilon_x$ (the direction along beating plane, and perpendicular to the beating plane) as shown in Fig.~\ref{Fig:persist}-b. If this ratio is small, it suggests amplitude in x-direction is much more than that in $z$-direction, and the trajectory is linear along $x$-axis. If this ratio is close to one, it suggests that amplitudes in x-direction and in $z$-direction are same, and the trajectory can be circular. 
For lower values of $\rho$, i.e. $\rho=0,0.1$,  amplitudes is close to one, therefore shape of trajectory is nearly circular. Further,
intermediate value of $\rho =0.25$ gives elliptical trajectory.  For $\rho=0.25$, $\epsilon_z/\epsilon_x$ grows from a small value to one, and eventually reaches to the transition towards the butterfly to elliptical trajectory. For the higher torsion ratio, i.e., $\rho>0.5$,  $\epsilon_{z}/\epsilon_{x}$ is very small suggesting motion is on the x-y plane only, this suggests end-monomer moves nearly along the x-axis. With $Pe$, $\epsilon_{z}/\epsilon_{x}$ shifts progressively towards the higher values. The trajectory of the end monomer may become circular again in the large compressive force limit. This also confirms that the transition between beating to rotation is not sudden, it changes continuously with increasing compressive force.  As Fig.~\ref{Fig:persist}-b suggests, transition from beating to rotation follows the path of butterfly to circular via elliptical trajectory.

\begin{figure}[h]
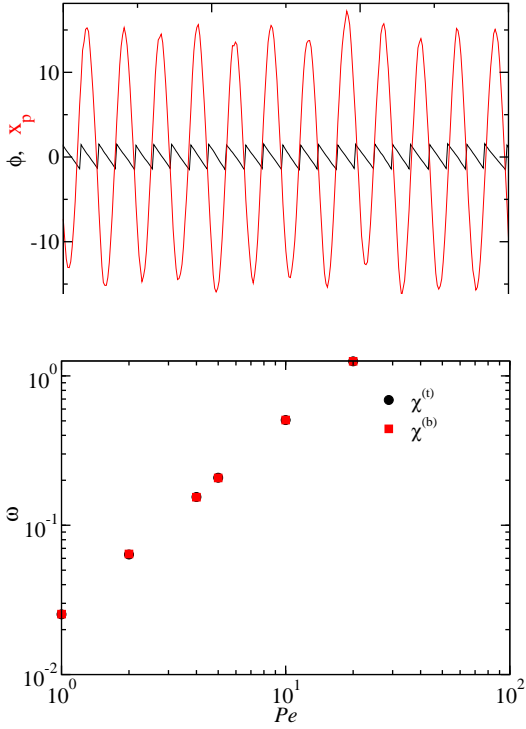

	\includegraphics[width=0.8\columnwidth]{periodicity}
	\includegraphics[width=0.8\columnwidth]{freq_energy}
	\caption{a) Time-series of the azimuthal angle and variation of x component of the terminus end for $Pe=10$ and $\rho=0$. b) Frequency of oscillation $\omega$ obtained from the bending ($\blacksquare$) and torsion  ($\bullet$) energies as a function of $Pe$.}
	\label{fig:period}
\end{figure}

We present here the discussion and the quantification of oscillation frequency through deflection of the ribbon in  x-direction.  The deflection is caused by bending of the filament, therefore measured frequency  in this way quantifies only bending oscillations. Further, we analyse the oscillation frequency via the azimuthal angle of the ribbon's terminus  end. This quantity can provide  out-of plane oscillations at small $\rho$. In this limit, i.e., at $\rho = 0$ the oscillation of azimuthal angle coincides with the oscillation in x-deflection as illustrated  in  Fig.~\ref{fig:period}-a. When $\rho\simeq 0$, we observe local oscillations in the torsional parameter $\chi_i^t$ as given in Fig. 6-c of the main text.  The frequency of oscillations in bending energy is computed by the Fourier transformation of the time evolution of $\chi^b(t)$ for different values of $Pe$. Similarly, we also obtain frequency from oscillation in torsional energy $\chi^t(t)$.   Frequencies from torsional and azimuthal's angle coincides with the bending frequency  Fig.~\ref{fig:period}-b. \\

%We provide here two movie files showing the dynamics of the clamped active filament.

%Movie SI-MOVIE-1: The movie shows the rotational motion of the filament at $Pe = 4$ for $\rho=0$.

%Movie SI-MOVIE-2: The movie shows the beating motion of the filament at $Pe = 4$ for $\rho=1$. 

\end{document}